\documentclass{iau} 
\usepackage{graphicx,natbib}

{\newif\ifnotend
\notendtrue
\def\veclist{ABCDEFGHIJKLMNOPQRSTUVWXYZabcdefghijklmnopqrstuvwxyz.}
\def\top#1#2.{#1}
\def\tail#1#2.{#2.}
\loop\expandafter\xdef\csname v\expandafter\top\veclist\endcsname%
{{\noexpand\bf\expandafter\top\veclist}}
\edef\veclist{\expandafter\tail\veclist}
\if\veclist.\notendfalse\fi\ifnotend\repeat}
\def\spose#1{\hbox to 0pt{#1\hss}}
\def\lta{\mathrel{\spose{\lower 3pt\hbox{$\mathchar"218$}}
     \raise 2.0pt\hbox{$\mathchar"13C$}}}

\title[{\it Gaia:} from proposal to GDR1] 
{ {\it Gaia:}  from proposal to GDR1}

\author[Gerry Gilmore]   
{Gerard Gilmore}

\affiliation{Institute of Astronomy, Madingley Road, Cambridge CB3 0HA, UK \\ email: {\tt
gil@ast.cam.ac.uk}}

\pubyear{2017}
\volume{330}  
\jname{Astrometry and Astrophysics in the Gaia sky}
\editors{Alejandra Recio-Blanco, Patrick de Laverny, Anthony Brown and, Timo
Prusti eds.}
\begin{document}

\maketitle

\begin{abstract}
In this concluding article I recall the early history of the Gaia mission, showing that the original science case and expectations of wide community interest in Gaia data have been met. The quarter-century long partnership involving some 1,000 scientists, engineers and managers in industry and academia is delivering a large, high-quality and unique data set which will underpin astrophysics across many sub-fields for years to come.
\end{abstract}

\firstsection 
\section{{\it Gaia}: origins of the ESA mission}

The {\it Gaia} mission concept evolved naturally from the successful Hipparcos space astrometric mission. Hipparcos was launched in 1989, and operated until March 1993. It delived reliable parallaxes and proper motions for 118,000 stars, and was supplemented by a star-mapper flux and position catalogue, Tycho2, of 2.5million stars, being a nearly complete sample to magnitude 11. Hipparcos established that absolute parallaxes could indeed be determined from space observations, following an original suggestion by Pierre Lacroute. A brief overview of the Hipparcos mission, from its proposal to its completion, is provided by \cite{Perryman2011}.

 Astrometry from space has unique advantages over ground-based observations. All-sky coverage is possible, removing the challenge, with risk of systematic errors, of cross-calibrating complementary hemispheric facilities. A relatively stable and temperature- and gravity-invariant operating environment is viable.  Even more importantly, absolute astrometry is possible. Narrow field astrometry, from ground or with a typically-designed telescope, such as the Hubble Space Telescope, measures differential parallaxes between all objects in its narrow field of view. In a small field of view every object has similar angular distance from the Sun-Earth baseline, and so has similar parallactic angle. All parallactic ellipses are aligned, providing no information on absolute scale.  One attempts to convert to absolute parallaxes either by modelling the distance (parallax) of distant stars in the field, or by comparing to zero-parallax extragalactic objects, with due compensation for their different energy distributions and/or image structure. The possibility of systematic errors is always present. Space astrometry introduces, through appropriate optical design, a large differential angle between stars which are separated by only a small angle on the detector. Thus precise small-angle measurement, together with knowledge of the large angle offset delivered by the spacecraft optical design, compares stars with very different but known parallax factors, and so allows absolute parallax measures. Hipparcos delivered the large angle with an optical system which projected fields separated on the sky by 58$^{\rm o}$ through a modulating grid onto a single-pixel image dissector scanner detector. Gaia has two separate telescopes, with angular separation (``basic angle") of 106$^{\rm o}$, delivering two 0.$^{ \rm o}$7 fields of view onto a single very large focal plane, made of an array of 106 CCDs. Manifestly, observing two widely separated fields of view simultaneously from the ground is more difficult, and inevitably involves different observing conditions for each line of sight, though radio-wavelength VLBI achieves precision. Space access is required in the optical.  During the latter stages of the Hipparcos mission space astrometry had become a proven technique, so opportunities for successor missions to deliver the very wide science case thus enabled were investigated.
\begin{figure}
\begin{center}
\includegraphics[width=.8\hsize]{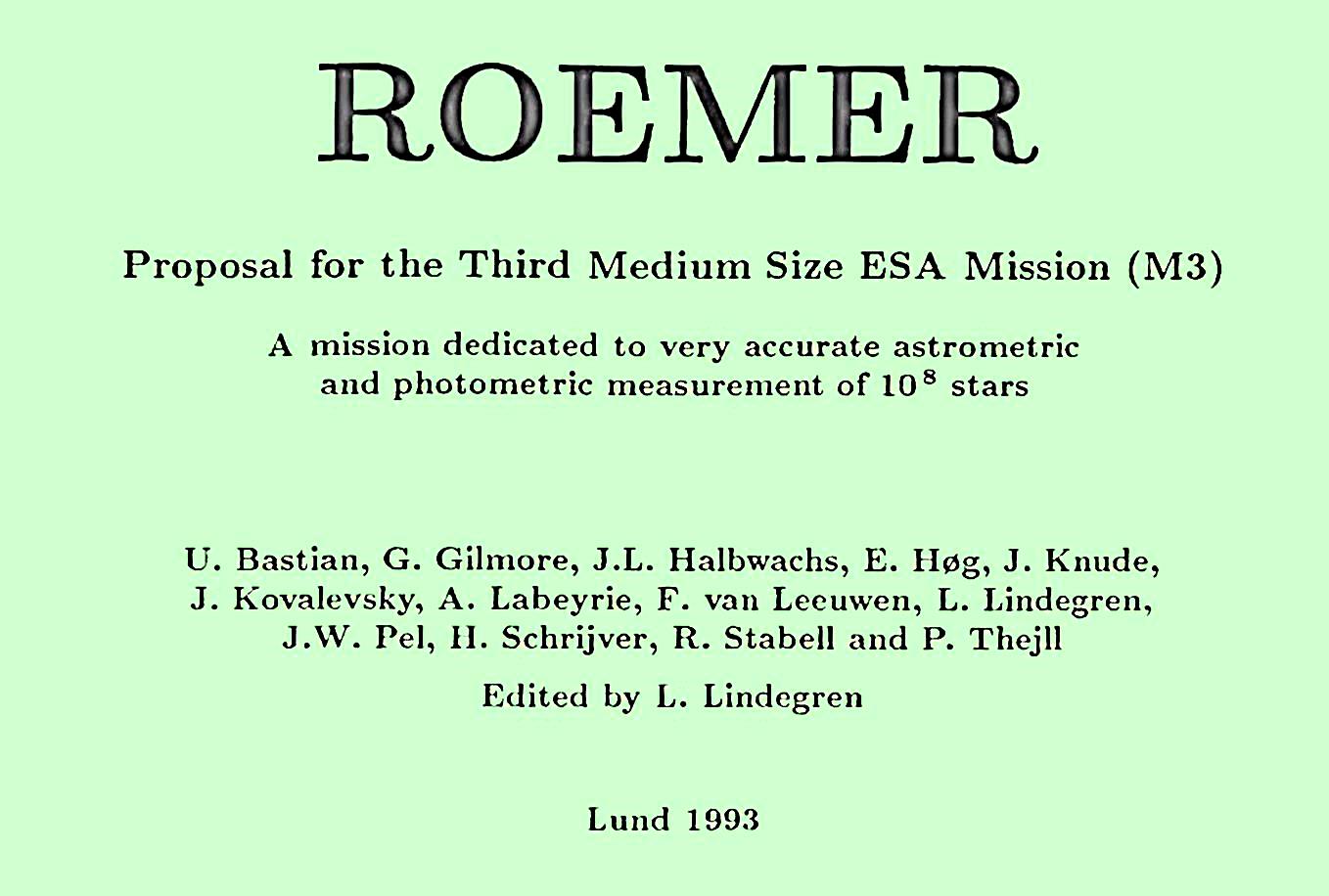}
\end{center}
\caption{The ROEMER proposal to ESA in 1993. The original is bright green.}\label{fig:Roemer}
\end{figure}

During the 1980's the scientific interest in Galactic Structure developed rapidly. The all-sky photographic surveys (Palomar POSS-II, ESO, UK Schmidt) were digitised leading to quantitative advances ranging from stellar luminosity functions, to discovery of the Galactic thick disk, to quantification of the local Dark Matter density, and to discovery of the Sgr dwarf galaxy and stellar streams in the Galactic halo, direct evidence that Galaxy evolution continued today, to name just some results of personal interest \citep{KTG93,GilmoreReid,KG2,IGI}.  This, together with rapid advances in understanding stellar evolution, provided the range of science cases and the large scientific community available to interpret and analyse the data which justified a much more ambitious determination of stellar distances, kinematics and chemical abundances.

 Technology had also advanced, with availability of large-format 2-D CCD detectors ensuring major efficiency and precision gains. The other aspect of the context was ESA's strategic interest in interferometry. Various strategic studies led to ESA reports SP-1135 {\it ``A Proposed Medium-Term Strategy for Optical Interferometry in Space"} and SP-354 {\it ``Targets for Space-Based Interferometry"}, proposing global astrometry as a high-priority area for space interferometry. 

\begin{figure}
\begin{center}
\includegraphics[width=.8\hsize]{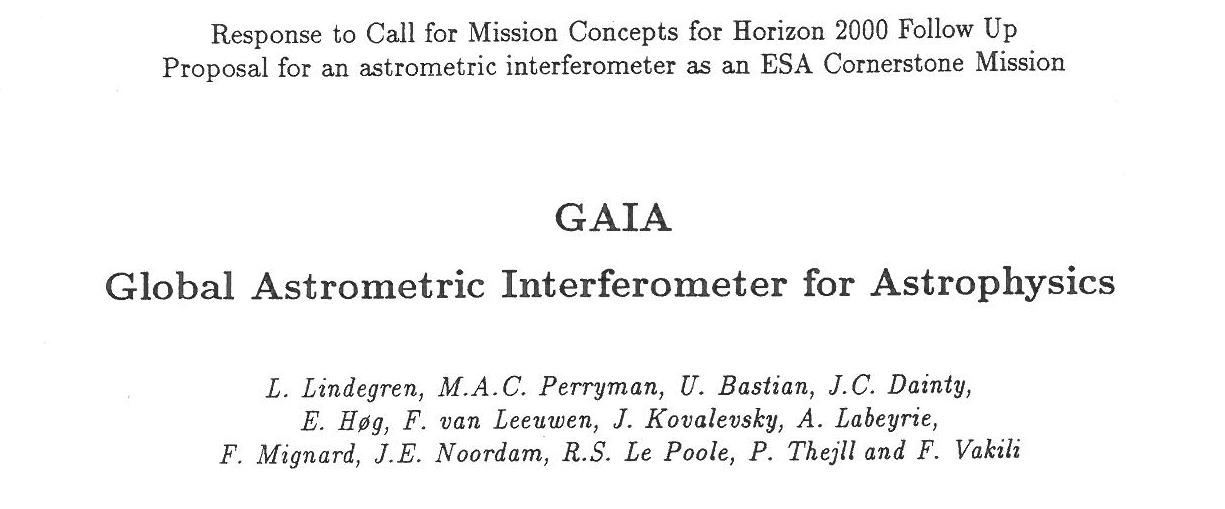}
\end{center}
\caption{The 1993 proposal for a Cornerstone interferometric astrometric mission.}\label{fig:Gaia}
\end{figure}

In this context two proposals  were submitted. The first (May 1993), to the ESA M3 Call (1993), was for the ROEMER concept (Fig.~\ref{fig:Roemer}), to provide astrometric and photometric data with 100microarcsec precision for $10^8$ stars, was highly rated. The second (October 1993) (Fig.~\ref{fig:Gaia}) which introduced the acronym GAIA - Global Astrometric Interferometer for Astrophysics - was more ambitious for a Cornerstone mission under the Horizon 2000 programme, to observe $5.10^7$ objects with accuracy 10microarcsec at magnitude 15. In an important Annex to the ROEMER proposal, J. Kovalevsky noted that use of CCDs additionally removed the need for an Input Source Catalogue, as Hipparcos required, so that the mission could be a true survey mission. These proposals led to an ESA-funded industrial study complemented by an astronomy Science Advisory Group. This group was tasked to develop the science case for a Cornerstone mission. The group was Chaired by the ESA Project Scientist, Michael Perryman, with Project Manager Oscar Pace, and as members G. Gilmore, E. Hoeg, M. Lattanzi, L. Lindegren, F. Mignard, S. Roeser, and P.T. de Zeeuw. Roeser was later replaced by K. de Boer, and X. Luri joined.

\begin{figure}[h]
\begin{center}
\includegraphics[width=.7\hsize]{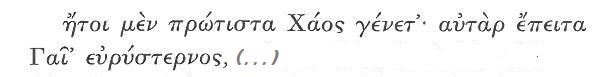}
\includegraphics[width=.7\hsize]{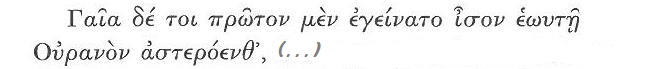}
\end{center}
\caption{{\it Gaia},
Hesiod, {\it Theogony  116/117 \& 126/127} Loeb Classical Library, Hesiod 1, 2006.}\label{fig:hesiod}
\end{figure}

Interestingly, rather early in the industry study it became apparent that interferometry was not the optimum technical solution to the scientific challenge. This forms yet another example of the common lesson that solutions looking for problems are rarely implemented. The initial acronym GAIA mutated into the name {\it Gaia} and survives. It turns out to be an appropriate name in its own right - and motivating the fairing logo (Fig~\ref{fig:people}) - after the ancient goddess as she appears, for example, in Hesiod's {\it Theogony 116/117 \& 126/127} (Fig.~\ref{fig:hesiod}) who came into being after {\it Chaos} and generated the starry sky. One interpretation of her coming into being is as a contrast to the unintelligible ({\it Chaos}, a gap, a wide opening) and as a generator of the explorable (the starry sky amongst many others).

The Study Team developed the {\it Gaia Concept and Technology Study Report} ESA-SCI(2000)4, commonly referred to as the mission ``Red Book", although the printed cover is white, which is the full proposal for the Gaia mission. [The full document remains available via the ESA {\it Gaia} web site, while an accessible summary was published in \cite{Perryman2001}.] Following the study the {\it Gaia} mission was formally presented to a selection meeting of the ESA communities, together with the other proposed missions, on September 13, 2000, in Paris. Presentations made were
\begin{itemize}
\item Scientific Case:     P.T. de Zeeuw
\item Payload, Accuracy and Data Analysis: L. Lindegren
\item Spacecraft and Mission Implementation:  O. Pace
\item Why, How and When?  G. Gilmore
\end{itemize}

Following these presentations {\it Gaia} was adopted as a Cornerstone Mission. Detailed spacecraft design and cost-reduction-motivated redesign continued under an industry prime contract to Astrium (now Airbus Defence \& Space), leading to the spacecraft successfully operating today.

\section{The {\it Gaia} science case - then and now}

The ``Key Science Objectives" presented for approval of the {\it Gaia} mission addressed the top-level ambition to provide the data needed to describe the origin, formation and evolution of the Galaxy.
\begin{itemize}
\item Structure and Kinematics of our Galaxy
\begin{itemize}\item shape and rotation of the bulge, disk and halo
\item internal motions of star forming regions, clusters, etc
\item nature of spiral arms and the stellar warp
\item space motions of all Galactic satellite systems
\end{itemize}
\item Stellar populations
\begin{itemize}
\item physical characteristics of all Galactic components
\item initial mass function, binaries, chemical evolution
\item star formation histories
\end{itemize}
\item Tests of Galaxy Formation
\begin{itemize}
\item dynamical determination of dark matter distribution
\item reconstruction of merger and accretion history
\end{itemize}
\end{itemize}

Other science products of the survey mission included Stellar Astrophysics, from luminosity calibration of large samples, including distance scale calibration; studies of the Solar System, with unique capability to map potentially earth-crossing asteroids orbiting interior to 1AU; discovery of large volume-complete samples of extra-solar planets;  important contributions to galaxies and quasars; and establishment of a dense high-precision reference frame.  Dynamical determination of the Galactic gravitational potential to determine the distribution of Dark Matter was a major goal. Although the term ``Dark Energy" was not in wide use at the time of the proposal, the science case included both precision distance calibration in cosmology and mapping orbits of nearby galaxies, from the Galactic halo into the Hubble flow. Particular emphasis was made on contributions to tests of General Relativity and the metric.This is an obvious science of interest for {\it Gaia}, since light-bending by the Sun (and planets) is the largest astrometric signal for most sources - Solar light bending is 4milliarcsec even at $90^{\rm o}$ from the Sun. Other GR effects include strong- and  micro-lensing, and possible detection of gravitational waves. Indeed a special ``experiment" is underway with {\it Gaia} to measure light-bending by Jupiter, a first.

The science case summary in the year 2000 was
\begin{itemize}
\item {\it Gaia} will determine
\begin{itemize}
\item when the stars in the Milky Way formed
\item when and how the Milky Way was assembled
\item how dark matter in the Milky Way is distributed
\end{itemize}
\item {\it Gaia} will also make substantial contributions to
\begin{itemize}
\item stellar astrophysics
\item Solar System studies
\item extra-solar planetary science
\item cosmology
\item fundamental physics
\end{itemize}
\end{itemize}

This remains a topical summary of research activity in 2017. In part this is recognition that data of the type and accuracy and volume which {\it Gaia} is designed to deliver cannot be obtained in any other way. {\it Gaia} remains at the forefront of the field.

\subsection{From science case to data analyses}

An important aspect of the {\it Gaia} proposal was to deliver early and regular data releases to ensure access to data by the wide community, in spite of the need to accumulate data over long times to enable precise astrometry. One special aspect of the proposal was what has become the Gaia transient science alerts (https://gaia.ac.uk). These are proving of both intrinsic scientific interest and an opportunity to involve the wider community, including amateur astronomers and schoolchildren, in the ground-based follow-up. More generally regular major data releases - GDR1 being just the first - are designed to get as much data published and  available for analysis as early as possible. This approach is complemented by a very special feature of the {\it Gaia} mission which was an integral aspect of the original proposal case: {\it Gaia} data are made available freely to the entire astronomical community as soon as possible. There is no proprietary time, or proprietary science, retained for private analysis inside the data processing consortium. This approach was recommended by the original Study Team mentioned above because of both the volume of the data, and because of the very wide range of science analyses and applications possible. The enthusiastic and wide involvement of the community in GDR1 data shows the wisdom of this philosophy.

\begin{figure}[h]
\begin{center}
\includegraphics[width=\hsize]{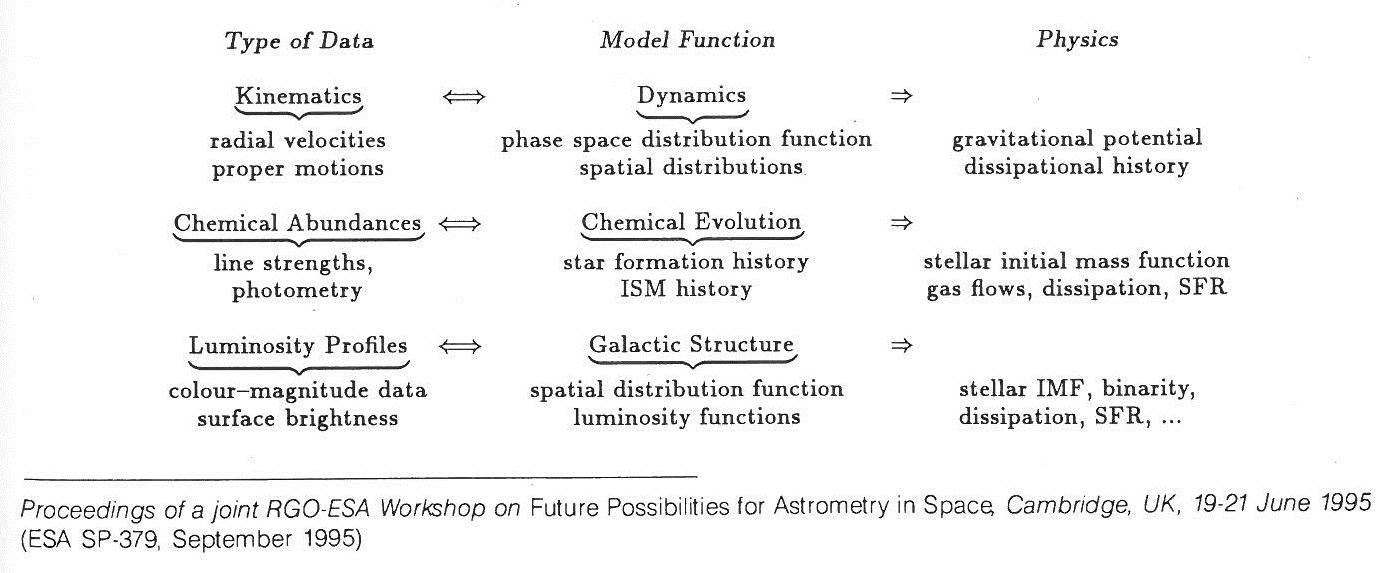}
\end{center}
\caption {A view of the approaches to data analysis from one of the {\it Gaia} preparation meetings. }\label{fig:models}
\end{figure}

Nonetheless, reliable intepretation of astrometric data is not a trivial task: the data sets are large, the errors are not simple, correlations are everywhere. Simply inverting or combining parallaxes is not the thing to do. Prior to {\it Gaia} astrometric data analyses were a specialist topic with rather few practitioners. Thus the {\it Gaia} data processing team has been concerned to ensure the wider community has the tools and expertise to analyse the data robustly. One example is the tutorial papar on how to estimate distances from parallaxes \citep{CABJ2015}. Another is the top-level analysis protocol methodology summarised in Fig.~\ref{fig:models}. The article by Binney in this Proceedings is an example of how the wider community has risen to the {\it Gaia} data challenge. Schools, tutorials, workshops, Challenges and so on are now widespread across the community.

\section{Lessons from history}

The GDR1 data release is the largest and highest accuracy and precision astrometric data set yet available. Almost daily papers using the data appear. It has certainly achieved its primary aim of introducing the astronomical community to large astrometric data sets. Nonetheless, GDR1 with its 2million TGAS parallaxes is tiny compared to future data releases. GDR2 will be 3 orders of magnitude larger, and start to make available the wealth of photometric, spectroscopic, spectrophotometric and time-series data, as well as derived quantities, for the more than 2.5billion objects which Gaia observes.

\begin{figure}[h]
\begin{center}
\includegraphics[width=\hsize]{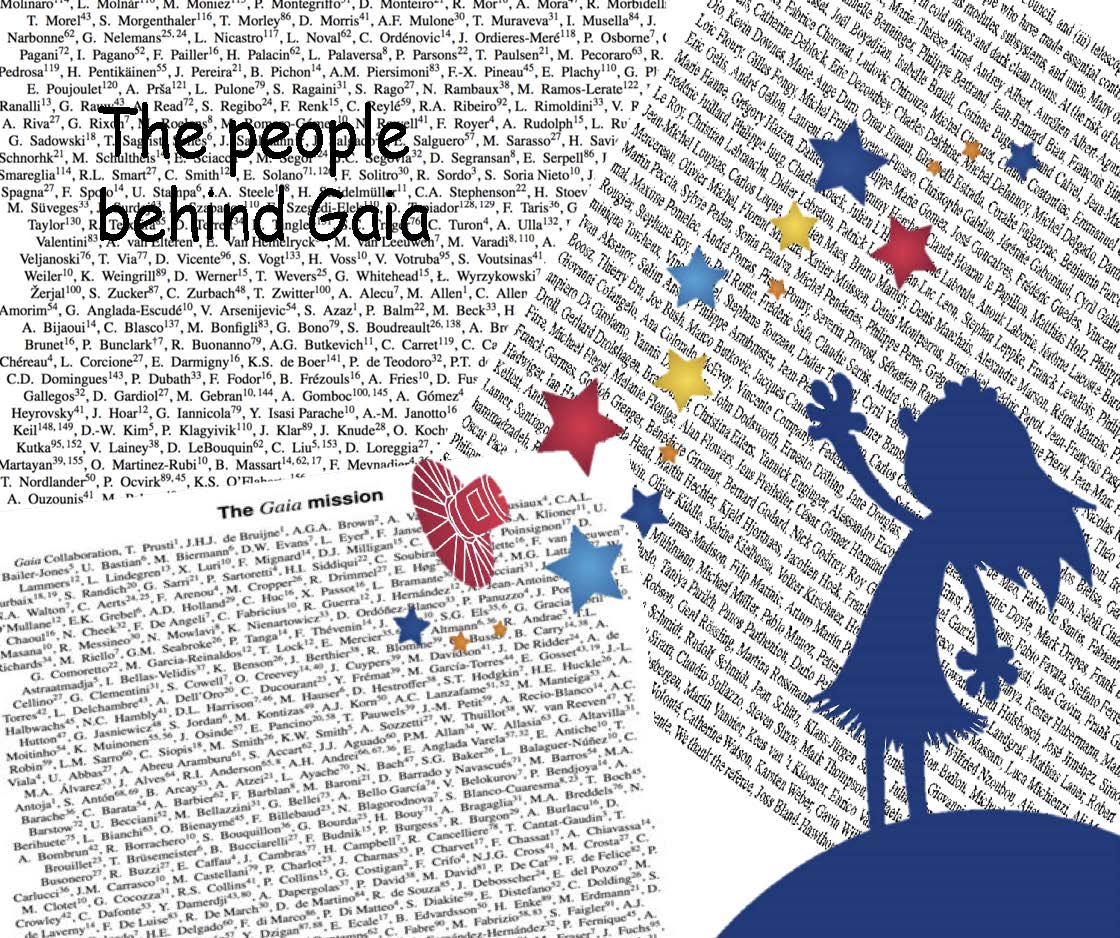}
\end{center}
\caption{The People behind {\it Gaia}, and the fairing logo. }\label{fig:people}
\end{figure}

Obtaining these data depends entirely on the superb {\it Gaia} spacecraft, designed, built and operated by very many talented and dedicated people in some 60 companies, under the leadership of the Prime contractor Astrium, now known as Airbus Defence \& Space, and soon to become simply Airbus. A large team at ESA and ESOC manage and implement the mission. These efforts are supported by the several hundred active participants in the   {\it Gaia} Data Processing and Analysis Consortium - DPAC. All these many people (Fig.~\ref{fig:people}) deserve our recognition and thanks. Over the quarter century of dedicated efforts since the first proposals for what became the {\it Gaia} mission and now, where we enjoy GDR1 and await GDR2 and its successors, there has been a close and productive partnership between ESA, the community and industry. We all benefit from that, and should acknowledge that we now stand on the shoulders of a giant team.

\section{Acknowledgements}
This work has made use of data from the European Space Agency (ESA) mission {\it Gaia} ({https://www.cosmos.esa.int/gaia}), processed by
the {\it Gaia} Data Processing and Analysis Consortium (DPAC, {https://www.cosmos.esa.int/web/gaia/dpac/consortium}). Funding for
the DPAC has been provided by national institutions, in particular the institutions participating in the {\it Gaia} Multilateral Agreement. 
Gilmore acknowledges partial support through the European Union FP7 programme grant ERC 320360, and the assistance of G. Pebody for Hesiod.

{}
\end{document}